\documentclass[12pt]{iopart}
\usepackage{graphicx}
\usepackage{aas_macros}

\usepackage{harvard}

        \newcommand{\secref}[1]{Section~\ref{#1}}
        \newcommand{\figref}[1]{Fig.~\ref{#1}}
        \newcommand{\eqnref}[1]{Eq.~\ref{#1}}

        \newcommand\optional[1]{}
        \newcommand{\Msun}{\ensuremath{\mathrm{M}_\odot}}

       \newcommand{\be}{\begin{equation}}
       \newcommand{\ee}{\end{equation}}
       \newcommand{\Ms}{\ensuremath{\mathrm{M}_\odot}}

\begin{document}

\title[Exploration of extrinsic parameters for gravitational-wave astronomy]{Physically motivated exploration of the extrinsic parameter space in ground-based gravitational-wave astronomy}

\author{V. Raymond$^{1,2}$ and W.M. Farr$^{2,3}$}

\address{$^{1}$ LIGO Laboratory, California Institute of Technology, MC 100-36, Pasadena CA, 91125,  USA} 
\address{$^{2}$ Dept.\ of Physics \& Astronomy and CIERA, Northwestern University, 2131 Tech Drive,  Evanston IL, 60208,  USA} \
\address{$^{3}$ School of Physics and Astronomy, University of Birmingham, Edgbaston, Birmingham B15 2TT, UK} 

\eads{\mailto{vivien@caltech.edu}, \mailto{w.farr@bham.ac.uk}}

\begin{abstract}

Efficient parameter estimation is critical for Gravitational-Wave
astronomy. In the case of compact binary coalescence, the high
dimensional parameter space demands efficient sampling
techniques---such as Markov chain Monte Carlo (MCMC). A number of
degeneracies effectively reduce the dimensionality of the parameter
space and, when known, can render sampling algorithms more efficient
with problem-specific improvements.  We present in this paper an
analytical description of a degeneracy involving the extrinsic
parameters of a compact binary coalescence gravitational-wave signal,
when data from a three detector network (such as Advanced LIGO/Virgo)
is available. We use this new formula to construct a jump proposal, a
framework for a generic sampler to take advantage of the
degeneracy. We show the gain in efficiency for a MCMC sampler in the
analysis of the gravitational-wave signal from a compact binary
coalescence.

\end{abstract}


\maketitle

\section{Introduction}

Among the sources of gravitational waves (GWs), inspiralling binary
systems of compact objects, neutron stars (NSs) and/or black holes
(BHs) in the mass range $\sim 1\,\Ms - 100\,\Ms$ stand out as likely
to be detected and relatively easy to model.  For the network of
ground-based laser interferometers \cite{Cutler:2001}, LIGO (Laser
Interferometer Gravitational-wave Observatory)
\cite{2009NJPh...11g3032A} and Virgo \cite{Acernese:2008b}, currently
undergoing upgrades, the detection-rate estimates for compact object
binaries, although uncertain, are expected to be about $70\,
\mathrm{yr}^{-1}$ \cite{ratesdoc}.

The detection of a gravitational-wave event is challenging and will be
a rewarding achievement by itself. After such a detection, measurement
of source properties holds major promise for improving our
astrophysical understanding of these sources and requires efficient
methods for parameter estimation.  This is a complicated problem
because of the large number of parameters ($15$ for spinning compact
objects in a quasi-circular orbit) and the quasi-degeneracies between
them \cite{Raymond:2009}, the significant amount of structure in the
parameter space, and the particularities of the detector noise.

We analyse the signal produced during the inspiral phase of two
compact objects of masses $M_{1,2}$ in quasi-circular orbit.  A
circular binary inspiral with both compact objects spinning is
described by a 15-dimensional parameter vector $\vec{\lambda}$. A
possible choice of independent parameters with respect to a fixed
geocentric coordinate system is:
\be
  \vec{\lambda} = \{m_1,m_2,d,t_c,\phi,\alpha,\delta,\iota,\psi, 
  a_\mathrm{spin1},\theta_\mathrm{spin1},\phi_\mathrm{spin1},a_\mathrm{spin2},\theta_\mathrm{spin2},\phi_\mathrm{spin2}\}
  \label{e:lambda}
\ee 
where $m_1$ and $m_2$ are the masses of the heaviest and lightest
members of the binary, respectively; $d$ is the luminosity distance to
the source; $\phi$ is an integration constant that specifies the
gravitational-wave phase at a reference frequency; the time of
coalescence $t_c$ is defined with respect to the centre of the Earth;
$\alpha$ (right ascension) and $\delta$ (declination) identify the source
position in the sky; $\iota$ defines the inclination of the binary
with respect to the line of sight; and $\psi$ is the polarisation
angle of the waveform.  The spins are specified by $0 \le
a_\mathrm{spin_{1,2}} \equiv S_{1,2}/M_{1,2}^2 \le 1$ as the
dimensionless spin magnitude, and the angles
$\theta_\mathrm{spin1,2}$, $\phi_\mathrm{spin1,2}$ for their
orientations with respect to the line-of-sight.

It is convenient to define two families of parameters. The intrinsic parameters:
\be
\overrightarrow{\lambda}_{intrinsic} = \{m_1,m_2, a_\mathrm{spin1},\theta_\mathrm{spin1},\phi_\mathrm{spin1},a_\mathrm{spin2},\theta_\mathrm{spin2},\phi_\mathrm{spin2}\} 
\ee
are required for the computation of the gravitational wave in any reference frame. The extrinsic parameters: 
\be
\overrightarrow{\lambda}_{extrinsic} = \{d,t_c,\phi,\alpha,\delta,\iota,\psi\} 
\ee
control the projection of the gravitational wave onto the geocentric reference frame, in which we can compute the response of each detector with \eqnref{eq:beam_pattern}.

Given a network comprising $n_\mathrm{det}$ detectors, we assume that the data collected at the $i-$th  
instrument ($i = 1,\dots, n_\mathrm{det}$) is given by $x_i(t) = n_i(t) + h_i(t;\vec{\lambda})$,  
where $h_i(t;\vec{\lambda})$ is the gravitational-wave signal (see \eqnref{eqn:signal}), and $n_i(t)$ is the detector noise (here assumed to be stationary and normally-distributed).

The equations governing the response of an observatory to gravitational waves have long been known, see for instance \cite{1973grav.book.....M} and references therein.
To illustrate the degeneracy present in this response we use Markov chain Monte Carlo (MCMC) methods to determine the multi-dimensional \emph{posterior} probability-density function (PDF) of the unknown parameter  
vector $\vec{\lambda}$ in equation~\ref{e:lambda}, given the data sets $x_i$ collected by a  
network of $n_\mathrm{det}$ detectors, a model $M$ of the waveform and the \emph{prior} $p(\vec{\lambda})$ on the  
parameters. 
One can compute the probability density via Bayes' theorem 
\be
p(\vec{\lambda}|x_j,M) = \frac{p(\vec{\lambda}|M) \, p(x_j|\vec{\lambda},M)}{p(x_j|M)}\,,
\label{e:jointPDF}
\ee
where 
\be
L \equiv p(x_j|\vec{\lambda},M) \propto 
\exp\left(
<x_j|h_j(\vec{\lambda})>-\frac{1}{2}<h_j(\vec{\lambda})|h_j(\vec{\lambda})>
\right)
\label{e:La}
\ee
is the \emph{likelihood function}, which measures the probability (under the noise distribution) of getting data $x_j$ given a signal $h_j$. 
The term $p(x_j|M)$ is the \emph{marginal likelihood} or \emph{evidence}. In the previous equation
\be
<x|y>=4Re\left( \int_{f_{\rm low}}^{f_{\rm high}}\frac{\tilde{x}(f)\tilde{y}^{*}(f)}{S_j(f)}\,\mathrm{d}f \right)
\label{e:prod}
\ee
is the \emph{overlap} of signals $x$ and $y$, $\tilde x(f)$ is the Fourier transform of $x(t)$, and $S_j(f)$ is the noise power-spectral density in detector $j$.    
The likelihood computed for the injection parameters $\mathcal{L}_\mathrm{inj}=p(x_j|\vec{\lambda}_\mathrm{inj},M)$ 
is then a random variable that depends on the particular noise realisation $n_j$ in the data $x_j=h(\vec{\lambda}_\mathrm{inj})+n_j$. 
The injection parameters are the parameters of the waveform template added to the noise.

To combine observations from a network of detectors with uncorrelated noise realisations 
we have the likelihood $p(\vec{x}|\vec{\lambda},M) = \prod_{a=1}^{n_\mathrm{det}}\, p(x_j|\vec{\lambda},M)\,$, for $\vec{x} \equiv \{x_j: j = 1,\dots,n_\mathrm{det}\}$ and
\be
p(\vec{\lambda}|\vec{x},M) = \frac{p(\vec{\lambda}|M)\, p(\vec{x}|\vec{\lambda},M)}{p(\vec{x}|M)}.
\label{e:Bayes}
\ee

The numerical computation of the PDF involves the evaluation of a large, multi-modal, multi-dimensional integral. MCMC 
methods \citeaffixed[and references therein]{gilks_etal_1996,GelmanCarlinSternRubin}{\emph{e.g.}}
have proved to be especially effective in tackling this numerical problem.

In the Markov chain Monte Carlo algorithm, a Markov chain crawls
around the parameter space according to a specific set of rules:
\begin{itemize}
\item{}At iteration n, the chain is in the state $\vec{\lambda}_n$. Choose a proposal state $\vec{\lambda}_k$ with probability $p(\vec{\lambda}_k|\vec{\lambda}_n)$.
\item{}Compute the acceptance probability $p_a$:
\be
p_a=\min\left\{ 1, \frac{p(\vec{\lambda}_k\,|\vec{x},M)p(\vec{\lambda}_n|\vec{\lambda}_k)}{p(\vec{\lambda}_{n}|\vec{x},M)\,p(\vec{\lambda}_k|\vec{\lambda}_n)} \right \} 
\ee
\item{}Accept $\vec{\lambda}_k=\vec{\lambda}_{n+1}$ as the new state of the chain with probability $p_a$, otherwise $\vec{\lambda}_{n+1}=\vec{\lambda}_{n}$ (with probability $1-p_a$)
\end{itemize}

The distribution of parameters in the set of states
$\left\{\vec{\lambda}_n\right\}$ of the chain following this procedure
converges towards the posterior distribution as $n \to \infty$. Note
that for any proposal to be included in this algorithm, the ratio
\be
r\equiv\frac{p(\vec{\lambda}_n |\vec{\lambda}_k)}{p(\vec{\lambda}_k | \vec{\lambda}_n)},
\ee
needs to be computed, see \secref{sec:detail_balance}.


We derive for the first time in the literature a proposal that generates jumps in parameter space that exploit a near-degeneracy in the detector responses for the three-detector case.  Using such a proposal in the context of an MCMC generates moves that efficiently explore the extrinsic dimensions of the posterior distribution function, even when the posterior is multi-modal with widely separated, narrow peaks in the extrinsic dimensions.

In this paper we first present the existing degeneracies involving the extrinsic parameters describing a binary coalescence in \secref{sec:deg}. In \secref{sec:eqn} we present the equations which we solve in \secref{sec:sol} to generate proposed moves. In \secref{sec:detail_balance} we address detailed balance. We apply our proposal in our Markov chain Monte Carlo algorithm and describe the results in \secref{sec:res}. Finally we conclude in \secref{sec:end}.


\section{Degeneracies between extrinsic parameters}
\label{sec:deg}


There exists a near-degeneracy in the detector response to a
gravitational wave involving the sky location (right ascension $\alpha$ 
and declination $\delta$), the polarization, $\psi$, the distance $d$ and the inclination $\iota$ of the source when
three non-collocated detectors are used. In the following discussion
we will restrict ourselves to the case of non-spinning signals for
simplicity.  Some of our approximations are inapplicable to spinning
signals, but we expect that our jump proposal may still prove useful
in the spinning case, particularly for signals that are weakly spinning.

\begin{figure} 
  \includegraphics[width=\linewidth]{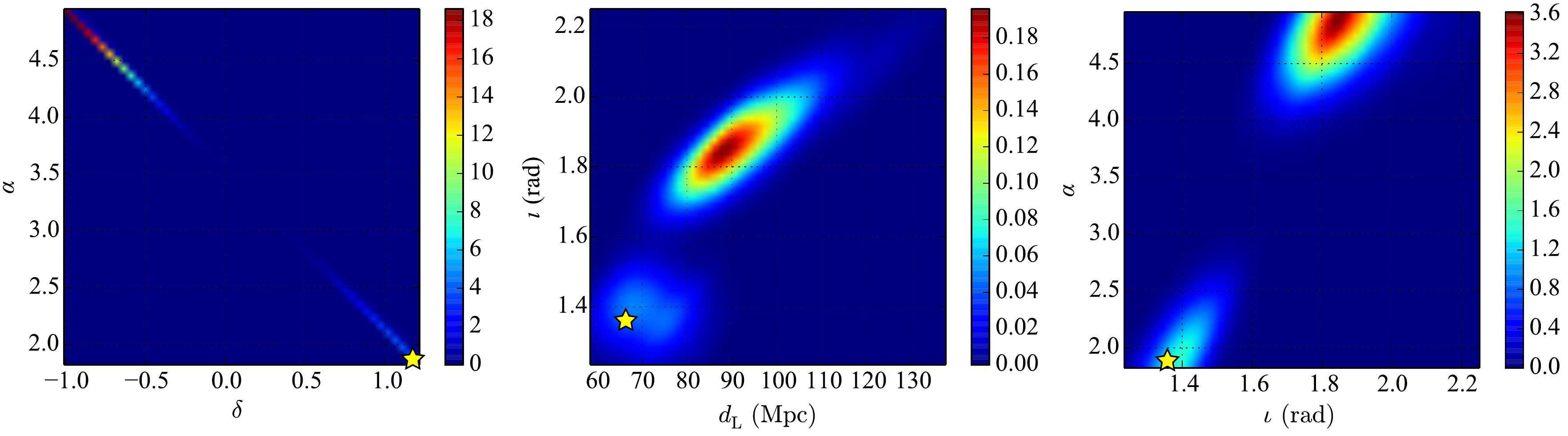}
  \caption{Simulation of a parameter recovery analysis. 
    The injected signal was a
    post-Newtonian non-spinning binary neutron star ($m_1=1.4\,\Msun$,
    $m_2=1.4\,\Msun$) at a signal-to-noise ratio of 20, recovered with
    a post-Newtonian frequency-domain non-spinning template model.
    Left: two dimensional probability density function in right ascension and declination. 
    The yellow star marks the injected values. 
    Center: two dimensional probability density function in inclination and distance. Each
    blob corresponds to one of the blobs in the left figure.
    Right: two dimensional probability density function in inclination and right ascension. Each
    blob corresponds to one of the blobs in the left figure.
    \label{fig:skymap_bimodal}
  }
 \end{figure}

The reflection of the true location of the source through the plane
defined by the three detectors conserves the arrival time at each
detector. This is the reason why in some three-detector analyses, two
modes in the sky location are recovered, see
\figref{fig:skymap_bimodal} (left).  The reflection condition
keeps the arrival time of the signal at each detector, $\Delta_1$,
$\Delta_2$ and $\Delta_3$, with $\Delta_j(\alpha,\delta,t_c)$, given
by:
\begin{equation}
\label{eq:time-of-arrival}
\Delta_j(\alpha,\delta)=S\cdot(-L_j)+t_c,
\end{equation}
constant.  Here the detector location is labelled by the vector $L_j$,
and the source by the vector $S(\alpha,\delta)$:
\begin{eqnarray}
S(\alpha,\delta) &= 
\left( \begin{array}{c}
 \cos\alpha\,\cos\delta \\
 -\sin\alpha\,\cos\delta \\
 \sin\delta 
\end{array} \right).
\end{eqnarray}

%
%
%
This degeneracy includes the time parameter $t_c$ as well, since the
reference time is at geocentre and the plane of the detectors does not
in general include the centre of the Earth.

This particular degeneracy has been well documented and a jump
proposal has been implemented involving the sky location and the
reference time, see for instance \cite{Veitch:2010}. However, the
detector network sensitivity pattern is not uniform on the sky. Any
change in sky location will change the effective strength of the model
template in each detector, and changes in polarization, inclination
and distance are needed to compensate. Both sky positions in
\figref{fig:skymap_bimodal} (left) correspond to different values
of polarization, inclination and distance. The center plot shows
the same blobs in the distance-inclination space, and the right plot shows 
the correlation between right ascension and distance. 


\section{Degeneracy Equations}

\subsection{Formulation of the equations}
\label{sec:eqn}

The signal in detector $j$, $h_j$, is the sum of two polarisations (in the non-spinning case):
\begin{eqnarray}
 h &= F_{j+}(\mathrm{HA},\delta,\psi)H_+(m_1,m_2,\iota,\phi,d,t_c) \nonumber \\
 &+F_{j\times}(\mathrm{HA},\delta,\psi)H_{\times}(m_1,m_2,\iota,\phi,d,t_c)
 \label{eqn:signal}
\end{eqnarray}

Where $F_{j+}$ and $F_{j\times}$ are the antenna beam patterns of the detector, relating the coordinate system centered on the detector to the coordinate system of the gravitational-wave source. $F_{j+,\times}(\mathrm{HA},\delta,\psi)$ are functions of the hour-angle $\mathrm{HA}$ (which is the right ascension $\alpha$ corrected for the earth's rotation: the Greenwich sidereal time minus the observatory's longitude and minus the right ascension), the declination $\delta$ and polarisation angle $\psi$ of the source. As a function of the right ascension, $F_{j+,\times}(\mathrm{HA},\delta,\psi) = F_{j+,\times}(\alpha,\delta,\psi;t_c)$. The antenna beam patterns are derived from the detector's three dimensional 2nd-order response tensor $D$ (which relates the local coordinates of the detector to the geocentric reference system where $\mathrm{HA}$, $\delta$ and $\psi$ are defined). For details and derivation, see \cite{creighton2012gravitational}.
\begin{eqnarray}
F_{j+}(\mathrm{HA},\delta,\psi) &= X^T(\mathrm{HA},\delta,\psi)\cdot D_j\cdot X(\mathrm{HA},\delta,\psi) \nonumber \\
&- Y^T(\mathrm{HA},\delta,\psi)\cdot D_j\cdot Y(\mathrm{HA},\delta,\psi) \\
F_{j\times}(\mathrm{HA},\delta,\psi) &= X^T(\mathrm{HA},\delta,\psi)\cdot D_j\cdot Y(\mathrm{HA},\delta,\psi) \nonumber \\
&+ Y^T(\mathrm{HA},\delta,\psi)\cdot D_j\cdot X(\mathrm{HA},\delta,\psi)
\label{eq:beam_pattern}
\end{eqnarray}
The vectors $X(\mathrm{HA},\delta,\psi)$ and $Y(\mathrm{HA},\delta,\psi)$ are:
\begin{eqnarray}
X(\mathrm{HA},\delta,\psi) &= 
\left( \begin{array}{c}
 -\cos\psi\,\sin\mathrm{HA} - \sin\psi\,\cos\mathrm{HA}\,\sin\delta \\
 -\cos\psi\,\cos\mathrm{HA} + \sin\psi\,\sin\mathrm{HA}\,\sin\delta \\
 \sin\psi\,\cos\delta 
\end{array} \right) \\
 Y(\mathrm{HA},\delta,\psi) &= 
\left( \begin{array}{c}
 \sin\psi\,\sin\mathrm{HA} - \cos\psi\,\cos\mathrm{HA}\,\sin\delta \\
 \sin\psi\,\cos\mathrm{HA} + \cos\psi\,\sin\mathrm{HA}\,\sin\delta \\
 \cos\psi\,\cos\delta 
\end{array} \right)
\end{eqnarray}

Thus, $F_{j+,\times}$ depend on the polarization angle, $\psi$ as 

\begin{eqnarray}
F_{j+}(\mathrm{HA},\delta,\psi) &= x_j(\mathrm{HA},\delta)\,\cos(2\psi)+y_j(\mathrm{HA},\delta)\,\sin(2\psi) \\
F_{j\times}(\mathrm{HA},\delta,\psi) &= y_j(\mathrm{HA},\delta)\,\cos(2\psi)-x_j(\mathrm{HA},\delta)\,\sin(2\psi),
\end{eqnarray}

or

\begin{eqnarray}
F_{j+}(\alpha,\delta,\psi;t_c) &= x_j(\alpha,\delta;t_c)\,\cos(2\psi)+y_j(\alpha,\delta;t_c)\,\sin(2\psi) \\
F_{j\times}(\alpha,\delta,\psi;t_c) &= y_j(\alpha,\delta;t_c)\,\cos(2\psi)-x_j(\alpha,\delta;t_c)\,\sin(2\psi),
\label{eqn:beam_pattern_psi}
\end{eqnarray}

where the functions $x_j(\alpha,\delta;t_c)$ and $y_j(\alpha,\delta;t_c)$ are complicated but known functions of the sky location.

The waveform polarisations $H_{+,\times}(m_1,m_2,\iota,\phi,d,t_c)$ are functions of the masses $m_{1,2}$, the inclination $\iota$ (angle between the line of sight and the orbital angular momentum), the phase at a reference time $\phi$, the distance to the observer $d$ and the time at coalescence $t_c$. And they can be written in the non-spinning case, considering only the dominant 2-2 mode ($H_+$ and $H_{\times}$ are then related by a simple $\frac{\pi}{2}$ phase shift), as:

\begin{eqnarray}
 H_+(m_1,m_2,\iota,\phi,d,t_c) &=-\frac{1+\cos^2(\iota)}{2\,d}H_+(m_1,m_2,\phi) \\
 H_{\times}(m_1,m_2,\iota,\phi,d,t_c) &=\frac{\cos\,\iota}{d}H_{\times}(m_1,m_2,\phi) = \frac{\cos\,\iota}{d}iH_+(m_1,m_2,\phi)
\end{eqnarray}

Abusing notation, from now on $H_{+,\times}$ refers to $H_{+,\times}(m_1,m_2,\phi)$, and we omit $t_c$, which simply provides an overall sliding of this component of the waveform (recall that $t_c$ also enters our analysis in Eq. \ref{eq:time-of-arrival}).
We define now two quantities of interest:

\begin{eqnarray}
 A_{j+}(\alpha,\delta,\psi,\iota,d;t_c) &=-\frac{1+\cos^2(\iota)}{2\,d}F_{j+}(\alpha,\delta,\psi;t_c) \\
 A_{j\times}(\alpha,\delta,\psi,\iota,d;t_c) &=\frac{\cos\,\iota}{d}F_{j\times}(\alpha,\delta,\psi;t_c)
 \label{eqn:a}
\end{eqnarray}

The signal amplitude is then given by:

\begin{eqnarray}
|| h || &= || A_{j+}(\alpha,\delta,\psi,\iota,d;t_c)H_++A_{j\times}(\alpha,\delta,\psi,\iota,d;t_c)H_{\times} || \\
 &= || A_{j+}(\alpha,\delta,\psi,\iota,d;t_c)H_++A_{j\times}(\alpha,\delta,\psi,\iota,d;t_c)iH_+ || \\
 &= || H_+ || \cdot || A_{j+}(\alpha,\delta,\psi,\iota,d;t_c) + iA_{j\times}(\alpha,\delta,\psi,\iota,d;t_c) || \\
 &= || H_+ || \sqrt{A_{j+}^2 + A_{j\times}^2}
\end{eqnarray}




To keep the same likelihood values under a change of parameters, we keep constant for each detector $j$ the quantity:
\begin{eqnarray}
R_j^2&=A_{j+}(\alpha,\delta,\psi,\iota,d;t_c)^2 + A_{j\times}(\alpha,\delta,\psi,\iota,d;t_c)^2 \\
 &= A_{j+}(\alpha^\prime,\delta^\prime,\psi^\prime,\iota^\prime,d^\prime;t_c^\prime)^2 + A_{j\times}(\alpha^\prime,\delta^\prime,\psi^\prime,\iota^\prime,d^\prime;t_c^\prime)^2
\label{eqn:const}
\end{eqnarray}
and the arrival time:
\begin{equation}
\Delta_j(\alpha,\delta,t_c)=\Delta_j(\alpha^\prime,\delta^\prime,t_c^\prime).
\label{eqn:det_time}
\end{equation}

This gives in the 3 detector network three additional constraints to the 3 arrival time constraints, and leads to a system of 6 equation and 6 variables. The solutions form a set of measure zero as expected, see for instance the narrow blobs (no lines nor extended surfaces) in \figref{fig:skymap_bimodal}. (The posterior distribution is composed of two blobs instead of two points because of the finite signal-to-noise ratio.)

\subsection{Solutions and proposal formula}
\label{sec:sol}

Starting from a set of parameters $\alpha, \delta, t_c, \psi, \iota, d$, we want to compute a new set  $\alpha^\prime, \delta^\prime, t_c^\prime, \psi^\prime, \iota^\prime, d^\prime$, which conserves \eqnref{eqn:const} and satisfies \eqnref{eqn:det_time}. We compute the quantities $R_j^2$ from \eqnref{eqn:const}. Using only \eqnref{eqn:det_time} for each of the three detectors gives the new values $\alpha^\prime, \delta^\prime, t_c^\prime$ from geometric arguments. The procedure consists of reflecting the sky position across the plane of the detectors and computing the corresponding $t_c$. This procedure is described in the literature, see for instance \cite{Veitch:2010} and references therein.

We now have the values of $\alpha^\prime, \delta^\prime, t_c^\prime$ and:
\begin{eqnarray}
F_{j+}(\alpha^\prime,\delta^\prime,\psi^\prime;t_c^\prime) &= x_j(\alpha^\prime,\delta^\prime;t_c^\prime)\,\cos(2\psi^\prime)& \nonumber \\
&+y_j(\alpha^\prime,\delta^\prime;t_c^\prime)\,\sin(2\psi^\prime) &=  F_{j+}^\prime(\psi^\prime)\\
F_{j\times}(\alpha^\prime,\delta^\prime,\psi^\prime;t_c^\prime) &= y_j(\alpha^\prime,\delta^\prime;t_c^\prime)\,\cos(2\psi^\prime)& \nonumber \\
&-x_j(\alpha^\prime,\delta^\prime;t_c^\prime)\,\sin(2\psi^\prime) &= F_{j\times}^\prime(\psi^\prime)
\end{eqnarray}


We can now write:
\begin{equation}
R_j^2=A_{j+}^2 + A_{j\times}^2=\left (-\frac{1+\cos^2(\iota^\prime)}{2\,d}F_{j+}^\prime(\psi^\prime)\right)^2 + \left(\frac{\cos\,\iota^\prime}{d}F_{j\times}^\prime(\psi^\prime) \right)^2
\end{equation}
We arbitrarily choose detectors $1$ and $2$ to write:
\begin{equation}
\frac{R_1^2}{R_2^2}=\frac{\left (1+\cos^2(\iota^\prime)\right)^2F_{1+}^\prime(\psi^\prime)^2 + 4\left(\cos\,\iota^\prime \right)^2 F_{1\times}^\prime(\psi^\prime)^2}{\left (1+\cos^2(\iota^\prime)\right)^2F_{2+}^\prime(\psi^\prime)^2 + 4\left(\cos\,\iota^\prime \right)^2 F_{2\times}^\prime(\psi^\prime)^2}
\label{eqn:R12}
\end{equation}
And detectors $2$ and $3$ to write:
\begin{equation}
\frac{R_2^2}{R_3^2}=\frac{\left (1+\cos^2(\iota^\prime)\right)^2F_{2+}^\prime(\psi^\prime)^2 + 4\left(\cos\,\iota^\prime \right)^2 F_{2\times}^\prime(\psi^\prime)^2}{\left (1+\cos^2(\iota^\prime)\right)^2F_{3+}^\prime(\psi^\prime)^2 + 4\left(\cos\,\iota^\prime \right)^2 F_{3\times}^\prime(\psi^\prime)^2}
\label{eqn:R23}
\end{equation}
\eqnref{eqn:R12} can be solved for $(cos\,\iota)^2$ to give:
\begin{eqnarray}
(\cos\,\iota^\prime)^2 &= \frac{R_1^2 \left(2 F_{2\times}^\prime(\psi^\prime)^2+F_{2+}^\prime(\psi^\prime)^2\right)-R_2^2 \left(2 F_{1\times}^\prime(\psi^\prime)^2+F_{1+}^\prime(\psi^\prime)^2\right)}{F_{1+}^\prime(\psi^\prime)^2 R_2^2-F_{2+}^\prime(\psi^\prime)^2R_1^2} \\
   & -2 \sqrt{\frac{\left(F_{2\times}^\prime(\psi^\prime)^2 R_1^2-F_{1\times}^\prime(\psi^\prime)^2 R_2^2\right)}{\left(F_{2+}^\prime(\psi^\prime)^2 R_1^2-F_{1+}^\prime(\psi^\prime)^2 R_2^2\right)^2}} \nonumber \\
   & \times\,\sqrt{\frac{\left(R_1^2 \left(F_{2\times}^\prime(\psi^\prime)^2+F_{2+}^\prime(\psi^\prime)^2\right)-R_2^2
   \left(F_{1\times}^\prime(\psi^\prime)^2+F_{1+}^\prime(\psi^\prime)^2\right)\right)}{\left(F_{2+}^\prime(\psi^\prime)^2 R_1^2-F_{1+}^\prime(\psi^\prime)^2 R_2^2\right)^2}} \nonumber
\label{eqn:iota}
\end{eqnarray}

Plugging this solution into \eqnref{eqn:R23} gives the equation for $\psi^\prime$, a linear equation of $\cos(4\psi^\prime)$ and $\sin(4\psi^\prime)$ 
\footnote{Some of the equations considered are very similar to the ones in \cite{1996CQGra..13.1279J}. However our equation \eqnref{eq:linear} differs from equation (48) in \cite{1996CQGra..13.1279J} which the authors find to be a polynomial equation of second order in the two variables $\cos(4\psi^\prime)$ and $\sin(4\psi^\prime)$ }:
\begin{eqnarray}
0 &=(R_3^2 x_2^2 y_1^2 - R_2^2 x_3^2 y_1^2 - R_3^2 x_1^2 y_2^2 + R_1^2 x_3^2 y_2^2 \nonumber \\
   & + R_2^2 x_1^2 y_3^2 - R_1^2 x_2^2 y_3^2) \cos(4\psi^\prime)  \\
   & + (-R_3^2 x_1 x_2^2 y_1 + R_2^2 x_1 x_3^2 y_1 + R_3^2 x_1^2 x_2 y_2 - R_1^2 x_2 x_3^2 y_2  \nonumber \\ 
   & + R_3^2 x_2 y_1^2 y_2 - R_3^2 x_1 y_1 y_2^2 - R_2^2 x_1^2 x_3 y_3 + R_1^2 x_2^2 x_3 y_3 \nonumber \\
   &  - R_2^2 x_3 y_1^2 y_3 + R_1^2 x_3 y_2^2 y_3 + R_2^2 x_1 y_1 y_3^2 - R_1^2 x_2 y_2 y_3^2) \sin(4\psi^\prime) \nonumber
\label{eq:linear}
\end{eqnarray}
Which we rewrite:
\be
0=a\cos(4\psi^\prime) + b\sin(4\psi^\prime)
\ee
The solution is then:
\begin{eqnarray}
\psi^\prime &= \frac{1}{2}\arctan\left(\frac{b-a\sqrt{\frac{a^2+b^2}{a^2}}}{a}\right) \nonumber \\
  & \textnormal{or} \nonumber \\
 \psi^\prime &= \frac{1}{2}\arctan\left(\frac{b+a\sqrt{\frac{a^2+b^2}{a^2}}}{a}\right)
 \label{eqn:psi_sol}
\end{eqnarray}

Only one of the two solutions of \eqnref{eqn:psi_sol} when plugged into \eqnref{eqn:iota} satisfy $ 0 \le \cos(\iota^\prime)^2 \le 1$. 

The distance $d^\prime$ can be computed using \eqnref{eqn:const} for any given detector:

\be
 d^\prime =\frac{\sqrt{\left(\frac{1+\cos^2(\iota^\prime)}{2}F_{j+}(\alpha^\prime,\delta^\prime,\psi^\prime) \right)^2 + \left(\cos\,(\iota^\prime)F_{j\times}(\alpha^\prime,\delta^\prime,\psi^\prime) \right)^2}}{R_j}
\ee

\subsection{Detailed balance considerations}
\label{sec:detail_balance}

For this proposal to be useful in a Metropolis-Hastings Markov chain Monte Carlo (as in \secref{sec:res}), one needs to compute the ratio of the probability densities on parameter space for particular jumps to be proposed:
\be
r\equiv\frac{p(\vec{\lambda} | \vec{\lambda}')}{p(\vec{\lambda}' | \vec{\lambda})},
\ee
where
\begin{equation}
\vec{\lambda}' = J(\vec{\lambda})
\end{equation}
is the point corresponding to $\vec{\lambda}$ under the mapping just described, which we denote by $J$, and 
\begin{equation}
p(x | y) 
\end{equation}
is the probability density for proposing point $x$ given that the current point is $y$.  In our case, the ratio of densities is given by 
\begin{equation}
\label{eq:proposal-ratio-exact}
r = \left| \frac{\partial J}{\partial \vec{\lambda}} \right|.
\end{equation}
Unfortunately, the function on parameter space described above is quite complicated, and its Jacobian even more so.  Rather than implementing the Jacobian directly, we use the following modified procedure for choosing a new parameter space point, $\vec{\lambda}'$ from $\vec{\lambda}$.  First, we compute 
\begin{equation}
\label{eq:modified-proposal}
\vec{\lambda}' = J(\vec{\lambda}) + \epsilon n',
\end{equation}
where $n$ is a randomly-chosen vector of $N(0,1)$ variates and $\epsilon$ is a scale factor that is much smaller that the dispersion we expect in the posterior about $\vec{\lambda}'$.  Let 
\begin{equation}
\label{eq:deviation-vector-definition}
\epsilon n = \vec{\lambda} - J^{-1}(\vec{\lambda}') \approx - \epsilon \frac{\partial J^{-1}}{\partial \vec{\lambda}'} \cdot n'.
\end{equation}
We do not need an analytic expression for the Jacobian to compute $n$---we only need to apply the mapping to $\vec{\lambda}'$ and subtract from $\vec{\lambda}$.  The proposal probability density ratio is given by 
\begin{equation}
\label{eq:proposal-ratio-numerical}
\frac{p(\vec{\lambda} | \vec{\lambda}')}{p(\vec{\lambda}' | \vec{\lambda})} = \frac{\phi(n)}{\phi(n')},
\end{equation}
where $\phi(x)$ is the PDF for the multivariate $N(0,1)$ distribution.  Based on the relation in \eqnref{eq:deviation-vector-definition}, \eqnref{eq:proposal-ratio-numerical} is consistent with \eqnref{eq:proposal-ratio-exact}, but we need not have an explicit expression for $\partial J/\partial \vec{\lambda}$.  Essentially, we have numerically computed the projection of the Jacobian on the $n'$ direction.  We use the modified proposal, \eqnref{eq:modified-proposal}, in what follows.

\section{Results from the jump proposal in a Markov chain Monte Carlo sampler}
\label{sec:res}

\begin{figure} 
  \includegraphics[width=\linewidth]{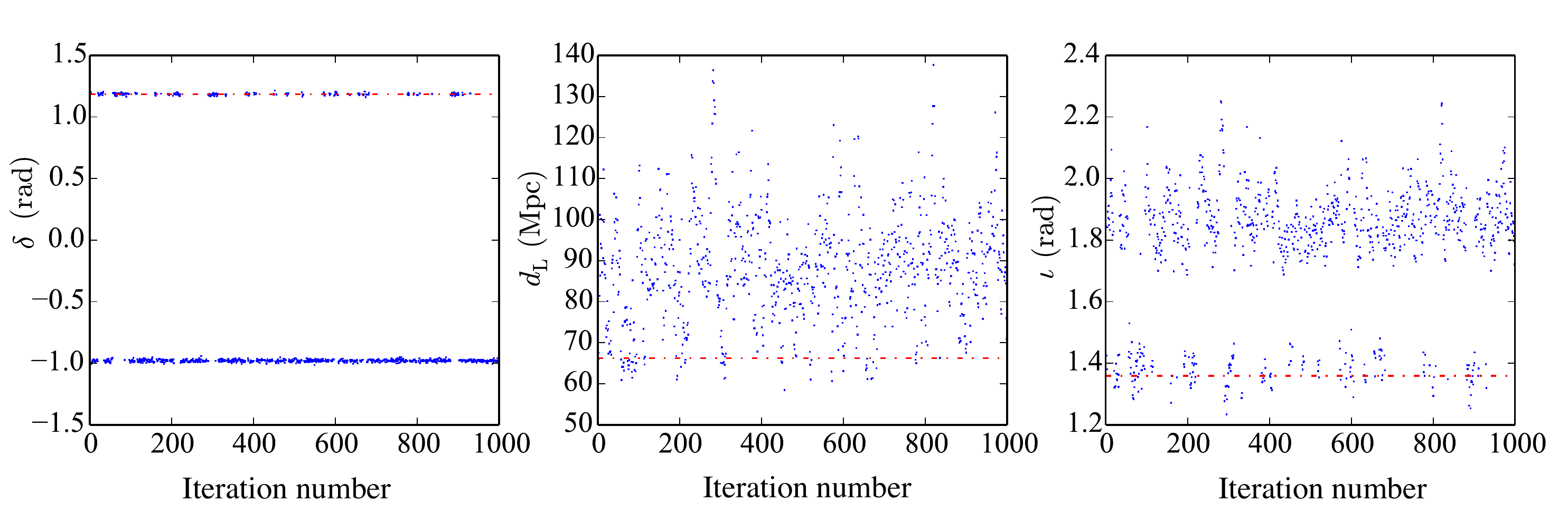}
  \caption{
   Plots of the samples from a Markov chain Monte-Carlo using our new proposal described in \secref{sec:eqn} as function of iteration number. The dot-dashed red line marks the injection value. Left: the declination parameter. Center: the distance parameter. Right: the inclination parameter. 
    \label{fig:new_prop}}
\end{figure}

\begin{figure} 
  \includegraphics[width=\linewidth]{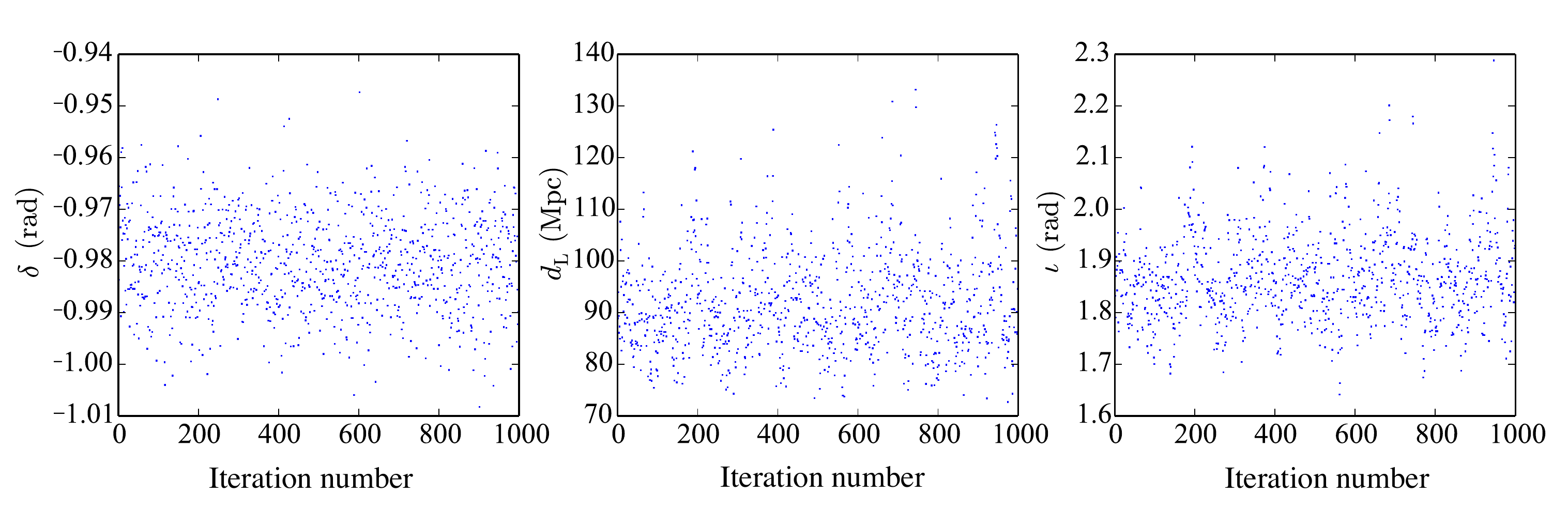}
  \caption{
   Plots of the samples from a Markov chain Monte-Carlo using the standard sky reflection proposal as function of iteration number. Left: the declination parameter. Center: the distance parameter. Right: the inclination parameter. 
    \label{fig:old_prop}}
\end{figure}

We have implemented the equations described in \secref{sec:eqn} as a
proposal in a Markov chain Monte Carlo sampling code. We present the
effect of including this proposal in \figref{fig:new_prop} and compare with 
the standard sky reflection proposal only in \figref{fig:old_prop}. 
We injected a known waveform from a non-spinning binary
neutron star system ($m_1=1.4\,\Msun$, $m_2=1.4\,\Msun$), computed
with post-Newtonian expansions \cite{Blanchet:2004ek}, into simulated
LIGO and Virgo noise at a signal-to-noise ratio of 20. 
The MCMC attempts to recover the posterior
density using the same frequency-domain template model and marginalising over the phase parameter \cite{margphi}.  In both
simulations we started in the reflected extrinsic parameter position
with respect to the true position. While the chain using the standard
sky reflection proposal \figref{fig:old_prop} gets stuck in the wrong mode, the
chain using our improved proposal \figref{fig:new_prop} finds the correct mode and samples both.

\section{Conclusions}
\label{sec:end}

We described in this paper a proposal which allows for a much better
exploration of the extrinsic parameter space for non-spinning
gravitational wave signals. It should still be helpful in the spinning
case, whose leading-order behavior mirrors the non-spinning case; we
plan to test this in future work.  It may be possible that using
an approximation beyond Quadrupole instead of \eqnref{eqn:const}
leads to a better handle on the spinning case where there is not
simple relation between $H_+$ and $H_{\times}$.  It may also be
necessary to include some intrinsic parameters to construct a more
efficient proposal for spinning analyses, as precession of the orbital
plane couples the spin parameters to the inclination.


\section*{References}
\bibliographystyle{jphysicsB}
\bibliography{extproposal.bib}

\end{document}